\documentclass[11pt]{article}

\usepackage{amsmath,amssymb,mathrsfs,upgreek}
\usepackage{graphicx}
\usepackage{float,placeins}
\usepackage{xcolor}
\usepackage{booktabs,multirow}
\usepackage{hyperref}
\usepackage[a4paper,margin=2.0cm]{geometry}
\providecommand{\bmhead}[1]{\section*{#1}}
\providecommand{\bibcommenthead}{}
\newcommand{\keywords}[1]{\par\noindent{\bf Keywords:} #1\par}
\newcommand{\beq}[1]{\begin{equation}\label{#1}\ignorespaces}
	\newcommand{\eeq}{\end{equation}}
\newcommand{\bea}[1]{\begin{eqnarray}\label{#1}\ignorespaces}
	\newcommand{\eea}{\end{eqnarray}}

\begin{document}
	
	\title{Merger Dynamics of N+N Co-planar Particles in Newton Gravitation}
	
	\author{
		Qi Su\thanks{Email: \href{mailto:SQphysics@emails.bjut.edu.cn}{SQphysics@emails.bjut.edu.cn}}
		\and
		Baicheng Zhang\thanks{Email: \href{mailto:bcZhangGR@emails.bjut.edu.cn}{bcZhangGR@emails.bjut.edu.cn}}
		\and
		Ding-Fang Zeng\thanks{Email: \href{mailto:dfzeng@bjut.edu.cn}{dfzeng@bjut.edu.cn}}
	}
	
	\date{
		Beijing University of Technology, Theoretical Physics Division,\\
		Beijing 100124, China
	}
	
	\maketitle
	
	\begin{center}
		{\small All signed authors contribute equally well to this work.}
	\end{center}

\abstract{We model the inspiral and merger dynamics of two co-planar rings in Newtonian mechanics with GR motivated corrections and illustrate their similarity with those of black hole binary systems on the orbital plane. Our simulation reveals a banana-shape deformation of the ``black holes'' involved, and a typhoon-like spiral structure in the merger product. Using an eXact One-Body approach, we compute the full gravitational waveform of this process and qualitatively reproduce results consistent with those of numerical relativity. Our simulation offers a transparent link between the feature of gravitational waveforms and the internal structure of black holes, thus a complementary interpretation of physics behind numerical relativity.}

	
\maketitle
	
\section{Introduction}\label{sec1}
Black holes (BHs), formed through gravitational collapse, are commonly considered to be singularities wrapped by event horizons. The spacetime curvature on the singularity diverges and all known physical laws lose predictability \cite{Hawking1972,Bejarano2017} there.  This leads to many challenges, such as the origin of Bekenstein-Hawking entropy and the information-loss puzzle et al. \cite{HawkingPenrose1970,Hawking1974,Almeida2021,Wallace2018,Danielsson1993,Hossenfelder2010}. To address these issues many fancy tools have been developed and applied, such as string theory and loop quantum gravity \cite{Witten1991,Assanioussi2020,Modesto2006}. These efforts suggest that BHs possess complex structures, from the resolved singularities to firewall type horizons \cite{Bojowald2020,Cadoni2022,Frolov2016,Almheiri2013,Mathur2014}. Some of them replace the event horizon with apparent horizon or other dynamic structures \cite{Nielsen2009,Murata2013,Abramowicz2002,Frolov2014,Thornburg1996,Schnetter2006,Centrella2010}. Although these attempts have yielded valuable insights, they usually obscure the simplicity of classical BHs, parameterised by mass $M$, spin $J$, and charge $Q$ exclusively.

Beyond these works, a complementary dynamical picture has been proposed and explored in Refs.~\cite{dfzeng2021,dfzeng2022,dfzeng2020,dfzeng2018a,dfzeng2018b,dfzeng2017,dfzeng2024}, where BHs are modeled as collapsars with an approximately and asymptotically $M(t,r)\approx\frac{r}{2G}$ mass-profile interior and a Schwarzschild-like exterior geometry. According to this picture, both the Schwarzschild and Lemaître time coordinate are regarded as legitimate time for the system's evolution description. This viewpoint motivates us to focus on the lasting-on, collapsar mass distribution and deformation, rather than the formation-finished-already, horizon and singularity. Such horizon and singularity do not exist \textcolor{red}{in} the time definition the probe is selected, is measured and is reported. 

A promising evidence for this proposal is the feature of gravitational waves (GWs) following from the merger of black hole binary (BHB) systems. In the Schwarzschild-time description, two BHs with successfully implemented horizon and singularity, would not fall into the horizon of each other within any finite Schwarzschild time definition of each other, due to the extreme gravitational time dilation effects. When two black holes get very close, isolating horizon of each other cannot be defined technically does not mean that the Schwarzschild time of each other would not be dilated. This renders it difficult to account for the transition from two initially point-like singularities to the final circular-line-like singularity characteristic of a Kerr BH, nor even the late-time damping feature. At the same time, while the effective-one-body (EOB) models \cite{Buonanno1999,Damour2007} can be calibrated to approximate QNM behavior, their analytical foundations alone often lack the ability to capture detailed late-time ringdown physics without NR input. In contrast, GW observations \cite{GW150914,Abbott2016GW2,Abbott2017GW170104,Abbott2017GW170814,Abbott2020GW190412,Abbott2020GW190521} have clearly detected waveforms featured by the late-time QNM like oscillations, indicating that the physical merger completes in finite time and the remnant BH indeed undergoes a ringdown phase.

Numerical relativity (NR) provides GR simulations of BHB mergers and predicts GW signals with late-time damping feature. In practice, NR tracks the dynamics of quasi-local horizons -- such as apparent horizon -- which deform strongly and can even undergo topology changes during the merger process \cite{Centrella2010,Toroidal99a,Toroidal99b,Toroidal16,Toroidal12}. These results reveal that the geometry of the merging objects on finite-time slices is highly nontrivial. However, the underlying simulations are technically involved and their geometric content is not always transparent at an intuitive level. This motivates the construction of simplified and phenomenological models that has potentials to capture the qualitative features of the process and its associated waveform morphology in a more visible and accessible way.

With this goal in mind, we investigate in this work the inspiral and merger dynamics of two co-planar rings in Newtonian gravity, and then supplement the dynamics with a GR-motivated time-dilation correction. We illustrate their similarity with those of BHB systems on the orbital plane. The framework of coupling Newtonian dynamics with the GW back-reaction has been well established in the pioneering works \cite{PoissonWill2014,Peters1963,Wagoner1976,Turner1977}.  
In these works, the GW energy loss was shown to be dominated by the variation of the mass quadrupole moment on the orbital plane. Our goal is to illustrate the similarity between the merger rings' structure and the apparent horizon of BHs of NR. The conservative part of the merger dynamics will be treated with the eXact One-Body (XOB) approach of \cite{dfzeng2024}. This approach allows us to explore the GW-like signal of the system within a computationally tractable framework.

The remainder of this paper is organized as follows. In Section II, we introduce the Newtonian framework and quadrupole radiation formula that form the basis of our merger model. Section III incorporates internal gravitation among particles within each ring, which will lead to asymmetric deformation of each single ring during inspiral. In Section IV, we include relativistic time-dilation effects to emulate gravitational time dilation near the horizon. Section V presents our computation of GW-like signal by the XOB method and their physical features. Finally, Section VI summarizes our results and discusses the implications for the interpretation of NR simulations and GW observations. In the numerical simulations we adopt units with $G=c=1$.

\section{Newtonian Mechanics and Quadrupole Radiation}

When considering the extended internal structure of BHs, their deformation will affect the quadrupole of the system thus the emission of GWs. Specifically, the outer side of the BHs in the orbit motion experience stronger energy loss comparing with the inner side because of their larger orbital radius and higher velocities of orbit motions. This uneven energy loss will lead to different accelerations within each BH, resulting in deformations. The outer regions compress more significantly than the inner regions, causing both BHs to exhibit a ``banana-shape deformation". As the merger progresses on, two ``banana-shaped"  BHs will keep stretching until their ends meet, becoming a single Kerr BH. In this work, we try to model the orbital plane feature of this deformation through a pair of co-planar rings. It should be emphasized that this ring model is intended as a heuristic framework to illustrate deformation dynamics, rather than as a physically rigorous description of the event-horizon of the true BHs.

As a whole, each ring maintains the same angular velocity at all points within the ring during the orbital motion process. Initially, we assume that the system is influenced only by gravitational radiation and neglect internal gravitational forces, aiming to preliminarily verify our conclusion regarding the "banana-shape deformation." 
We first define two circular rings with mass $M$, each consisting of $n$ particles of equal mass. The initial radius of the circular ring is $a_0=2GM/c^2$, and the centers of mass of the two rings are separated by a distance of $2r_0$. 
The initial coordinates of each particle in the system are represented as:
\begin{equation}
\left(r_0+a_0\cos{\frac{2\pi k}{n}},a_0\sin\frac{2\pi k}{n}\right),~k=1,2,\ldots,n, \label{ini.a} 
\end{equation}
\begin{equation}
\left(-r_0-a_0\cos{\frac{2\pi k}{n}},-a_0\sin\frac{2\pi k}{n}\right),~k=n+1,n+2,\ldots,2n. \label{ini.b} 
\end{equation}

In our toy model, the GW back-reaction is applied locally to each particle, not as a global damping term. Each mass point independently loses energy through its own quadrupole emission, and this local, point wise energy loss drives the non-uniform contraction and subsequent deformation. For the equal-mass configuration considered here, the system is centrosymmetric, and each diametrically opposite particle pair
contributes equally to the total quadrupole radiation, so pairwise summation correctly represents the global energy loss. In contrast, for unequal-mass binaries, one should evaluate the total quadrupole moment $
Q_{ij} = \sum_k m_k (x^k_k x^k_j - \tfrac{1}{3}r_k^2 \delta_{ij})$ in the center-of-mass frame, and then distribute the corresponding GW back-reaction to each particle according to its local coupling with $\dddot{Q}_{ij}$, which would require a more complicated mathematical scheme. Under this local GW back-reaction formula \cite{Peters1963}, the energy conservation for any pair of coordinate-symmetric particles reads:
\begin{equation}
-\frac{Gm^2}{2R_k (t)}-\frac{128G}{5c^5}m^2R^4_k (t) \omega^6\mathrm{d}t=-\frac{Gm^2}{2R_k (t+\mathrm{d}t)}, \label{con.E}
\end{equation}
where $R_k$ is the orbital radius of the particle pair $k$, $\omega$ is the angular velocity of the particle, the second term on the left hand side is the radiation back-reaction effects. In this model, we set the initial angular velocity of the rings as 
\begin{equation}
\omega_0=\sqrt{\frac{GM}{4r_0^3}}, 
\label{KeplerFormula}
\end{equation}
to maintain the system's stable circular motion. All particles have the same angular velocity with the rings. As the system evolves, $\omega$ depends on the orbital radius $r$ of the center of mass of the ring. Additionally, during the whole inspiral and merger process, we will fix the total linear momentum of the system to zero, so that no kick-off or memory effects will be considered in our model.

We introduce a dimensionless parameter $K$ to measure the effects of radiation back-reaction on the system at the initial time:
\begin{equation}
K=\frac{F_\mathrm{GW.back.reaction}}{F_\mathrm{mutual.gravity}}|_{t=t_0}. \label{def.K}
\end{equation}

It is defined as the ratio between the initial strength of the GW back–reaction and the mutual Newtonian attraction between the two rings. Within our Newtonian toy model, $K$ plays the role of a control parameter: for fixed geometrical setup, larger $K$ corresponds to stronger radiation reaction relative to mutual gravity and thus to a faster rate of deformation, while smaller $K$ leads to slower morphological evolution. In particular, once the dimensionless parameter $K$ is fixed, rescalings of the overall mass and length scales that leave $K$ unchanged produce the same dimensionless dynamical behavior. In this sense, $K$ organizes families of simulations in our toy framework, but it is not intended to provide a direct mapping to realistic GR binaries.

To compute $K$, we express the GW back-reaction force as:
\beq{}
F_{\mathrm{GW.back.reaction}}=\frac{128G}{5c^5}M^2r_0^3\omega_0^5=\frac{4 G^{7/2} M^{9/2}}{5c^5 r_0^{9/2}}. 
\label{def.Fra}
\eeq
Similarly, the initial mutual gravitational force between the two rings is given by:
\begin{equation}
F_{\mathrm{mutual.gravity}}=\frac{GM^2}{(2r_0)^2}. \label{def.Fmu}
\end{equation}
Thus, the dimensionless parameter $K$ is explicitly formulated as:
\begin{equation}
K=\frac{16}{5}\cdot\big(\frac{GM}{c^2r_0}\big)^{5/2}=\frac{2 \sqrt{2}
}{5}\big(\frac{r_0}{a_0}\big)^{-5/2}. \label{exp.K}
\end{equation}
showing that it arises as a negative power of the ratio $r_0/a_0$ between the initial separation and the ring radius and therefore parametrizes how quickly the system departs from its initial nearly circular configuration.

\begin{figure}[ht]
\includegraphics[width=0.33\textwidth]{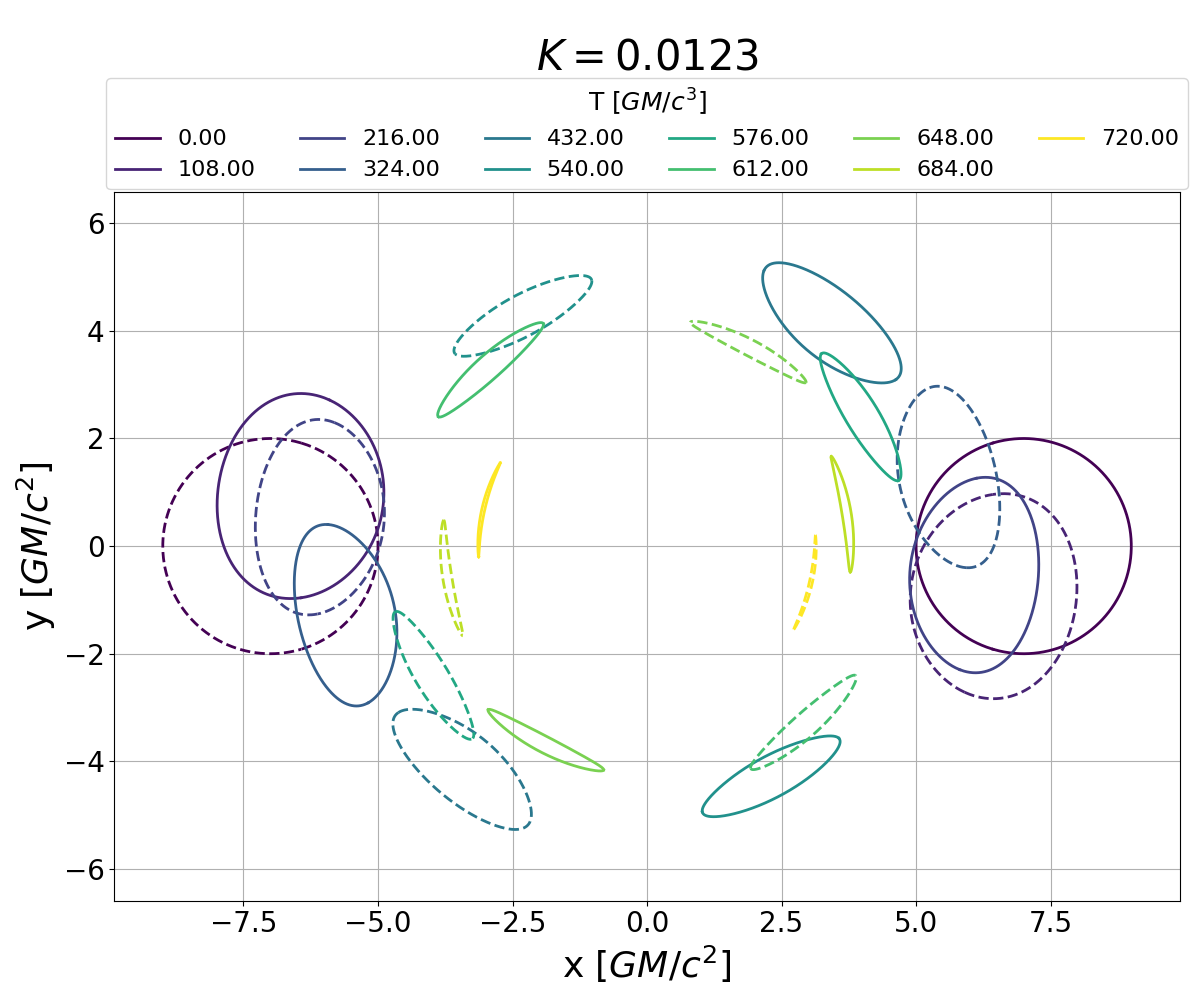} 
\includegraphics[width=0.33\textwidth]{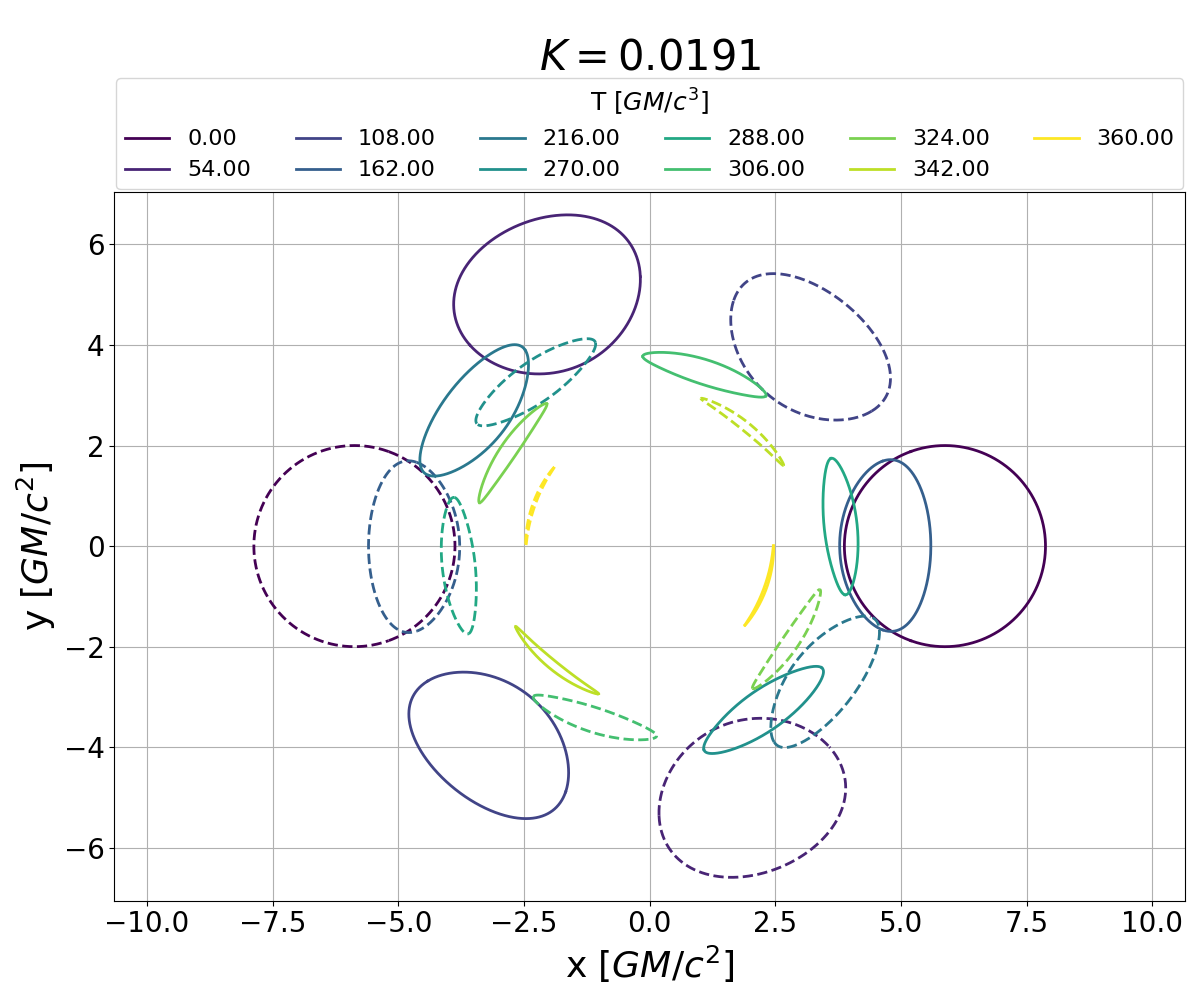} 
\includegraphics[width=0.33\textwidth]{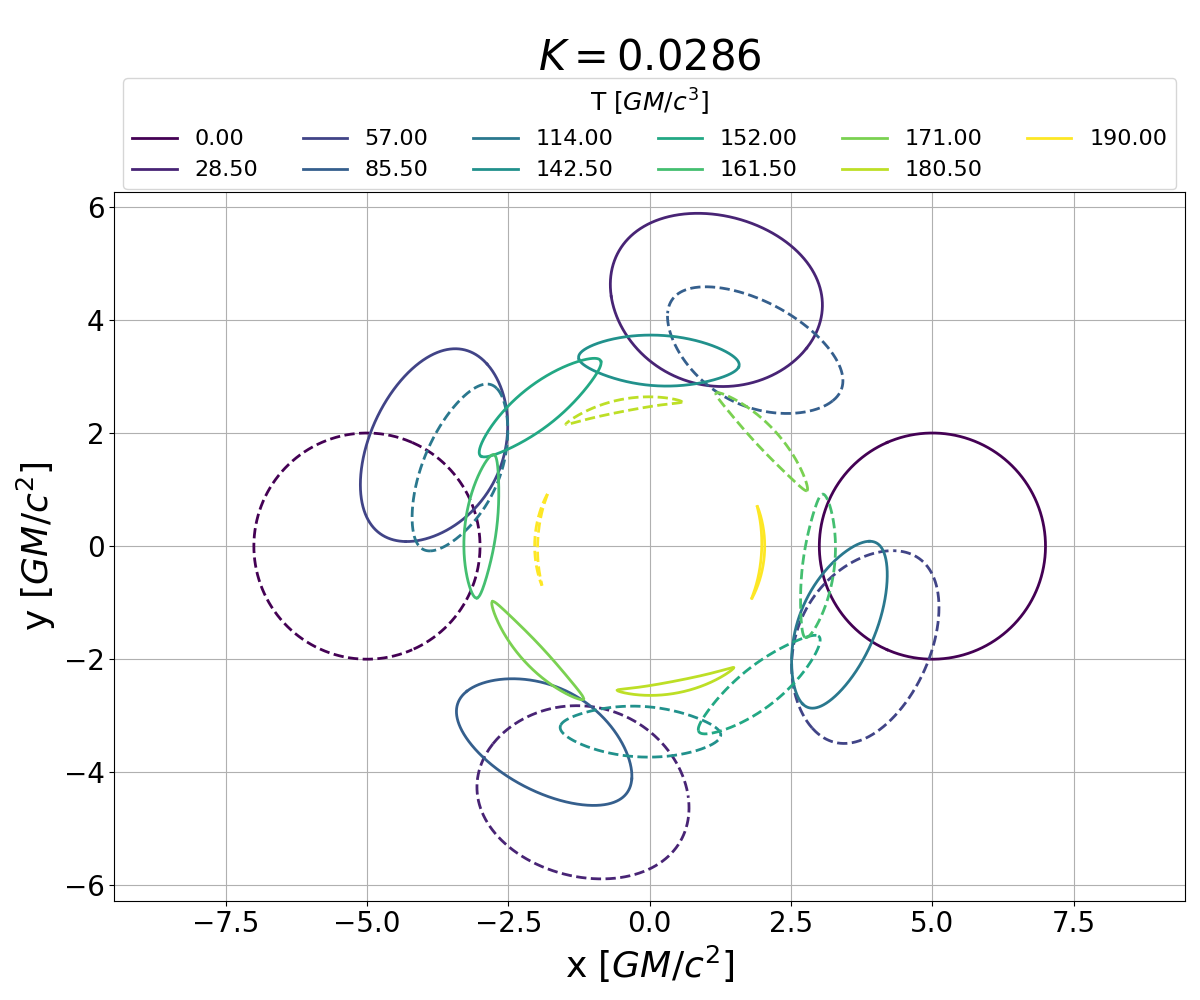} 
\caption{Evolution of the ring structures during binary merger under varying values of the parameter $K$. A higher $K$ indicates stronger GW back-reaction, leading to faster orbital decay and more pronounced deformation.}  
\end{figure}

For the event GW150914 \cite{GW150914}, $\omega_0=35$Hz, $m_1+m_2=65M_\odot$. Translating into our numerics, this corresponds to $r_0=2.922\frac{G(m_1+m_2)}{c^2}$ according to the Kepler formula \eqref{KeplerFormula}, so that $K=0.0191$. To align our numerical simulations with observation, we take $K=0.0123, 0.0191, 0.0286$, which corresponds to $r_0/a_0=3.50,2.94,2.50$. We displayed our numeric results in Fig.1. The time evolution in this figure is obtained through an adaptive Runge–Kutta (4/5) method with step-size control. The maximum residual error of all time points is $6.593\times10^{-4}$.

From the figure, we see that for larger $K$ values, the GW back-reaction dominates; this will lead to rapid orbital decay. Due to the uniform overall angular velocity, the outer orbit has a larger radius and thus more radiation power than the inner orbit. As a result, the outer rim undergoes stronger contraction, initiating a deformation of the ring structure. As $K$ decreases, mutual gravitational attraction becomes more significant, alleviating radiation-driven decay, so the system requires a longer time to exhibit a noticeable deformation.

These results confirm that $K$ is the fundamental parameter governing the initial conditions of equal-mass BH mergers with extended mass distributions. This also further confirms the viewpoint that each ring will experience banana-shape deformation in the inspiral stage.

\section{Internal Structure and Interaction of the Binary Merger System}
\label{secInternalStructure}

In the previous simulations, a uniform angular velocity was imposed across all particles within each ring, an approximation which neglects the effects of internal self-gravity. This approximation is overly restrictive. 
To refine this status and explore deformation dynamics in a more realistic setting, we extend our model by including effects of the gravitation from different parts of the ring internally. 
In this case, the gravitational force exerted on each mass point consisting of the ring can be written as:

\begin{equation}
{\vec{F}}_i=\sum_{j\neq i}^{2n} {Gm^2\frac{\boldsymbol{r}_i-\boldsymbol{r}_j}{\left|\boldsymbol{r}_i-\boldsymbol{r}_j\right|^3}}, \label{GravF}
\end{equation}
where $\boldsymbol{r}_i$ is the position vector of the $i$-th mass point.
By the same mechanism of local GW back-reaction introduced in the previous section,
each particle on the ring independently loses energy and angular momentum through its own quadrupole emission. The total energy and angular momentum conservation laws read,
\begin{equation}
-\sum_{j}\frac{Gm^2}{\left|\boldsymbol{r}_i-\boldsymbol{r}_j\right|}
+\frac{1}{2}m\left|\boldsymbol{v}_i\right|^2
-\frac{128G}{5c^5}m^2\left|\boldsymbol{r}_i\right|^4\omega_i^6\mathrm{d}t
=-\sum_{j}{\frac{Gm^2}{\left|\boldsymbol{r}_i^\prime-\boldsymbol{r}_j^\prime\right|}
+\frac{1}{2}m\left|\boldsymbol{v}_i^\prime\right|^2}, 
\label{Econ}
\end{equation} 
\begin{equation}
\sum_{i}m\left|\boldsymbol{r}_i\times\boldsymbol{v}_i\right|
-\frac{128G}{5c^5}\sum_{i}{m^2\left|\boldsymbol{r}_i\right|^5\omega_i^5\mathrm{d}t} 
=\sum_{i}m\left|\boldsymbol{r}_i^\prime\times\boldsymbol{v}_i^\prime\right|,
\label{Lconn}
\end{equation}  
where $\boldsymbol{r}_i^\prime$ and $\boldsymbol{v}_i^\prime$ represent the position and velocity of the $i$-th mass point after a time interval $\mathrm{d}t$, and $\omega_i$ is its effective orbital frequency.

In this case, it seems that a new dimensionless parameter $L$ appears. It quantifies the relative strength of the internal gravity compared with the mutual gravity.
\begin{equation}
\{K=\frac{F_\mathrm{GW.back.reaction}}{F_\mathrm{mutual.gravity}}, L = \frac{F_{\mathrm{internal.gravity}}}{F_{\mathrm{mutual.gravity}}} \}|_{t=t_0}.
\label{def.KL}
\end{equation}
where $F_{\mathrm{internal.gravity}}$ is defined from the self-gravitational potential energy $U_{\mathrm{self}}=GM^2/a_0$ of a ring:
\begin{equation}
F_{\mathrm{internal.gravity}} = \frac{GM^2}{a_0^2}, \label{def.Fin}
\end{equation}
Thus, the dimensionless parameter $L$ is explicitly formulated as:
\begin{equation}
\{K=\frac{2 \sqrt{2}}{5}\big(\frac{r_0}{a_0}\big)^{-5/2}
, L = 4\big(\frac{r_0}{a_0}\big)^{2}\}. \label{exp.L}
\end{equation}

Both $K$ and $L$ are toy control parameters used only to organize the qualitative behavior of the Newtonian ring model. In real GR binaries, radiation reaction, collapse timescales and internal gravity depend on curvature, redshift and time-slicing choices. So their numerical values here do not reflect those of real astrophysical systems.

From this expression, it is evident that $L$ explicitly depends on $r_0/a_0$, indicating that this ratio directly affects the system's evolution. In the following simulations, we will investigate the evolution of the system under different values of K and L, aiming to explore their effects on the final deformation patterns within our model.

\begin{figure}[ht]
\centering
\includegraphics[width=1.0\textwidth]{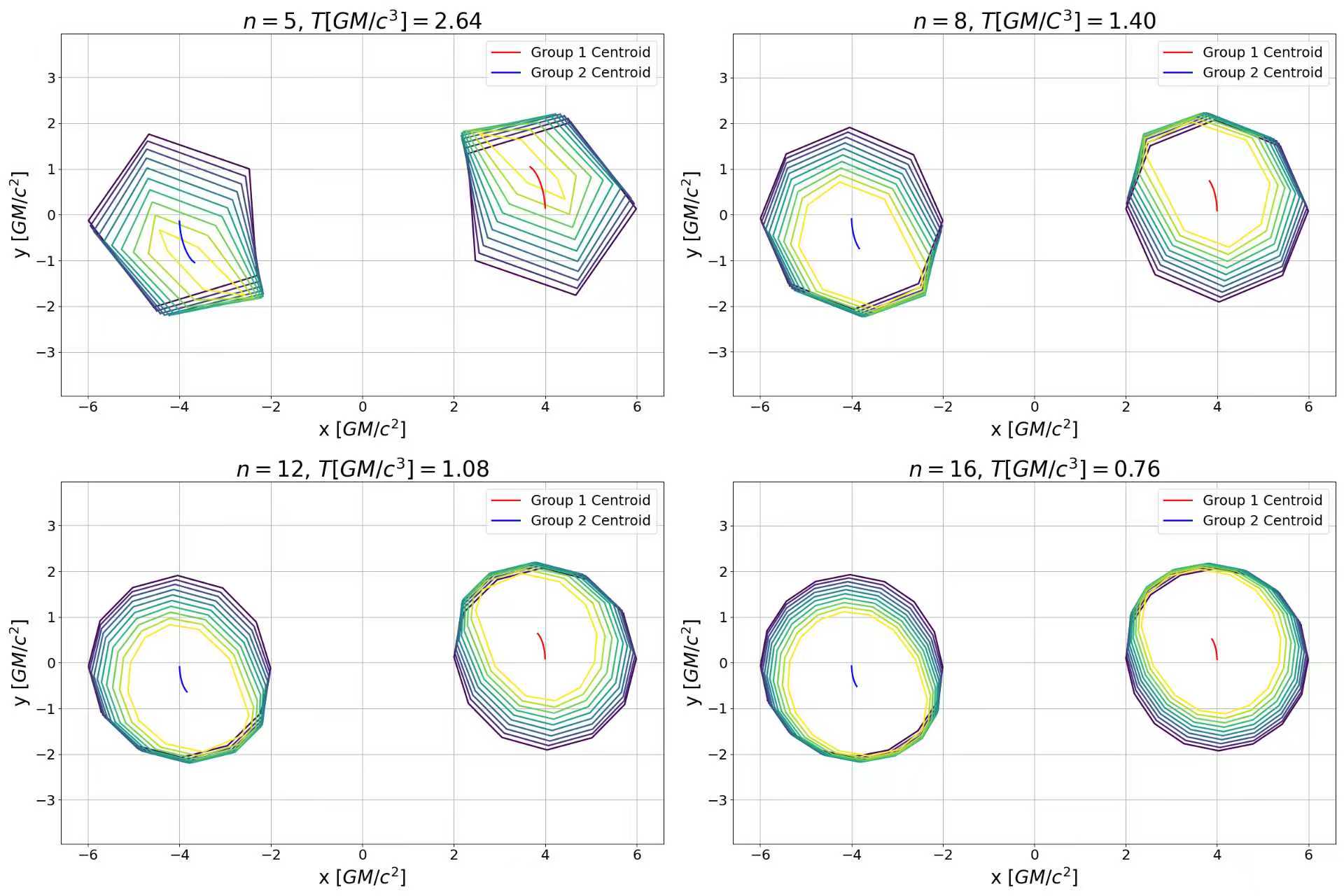}  
\caption{Influence of particle number $n$ on the deformation of ring profile. As $n$ increases, the system evolves with greater spatial resolution but reduced numerical stability, shortening the duration of coherent evolution. The trend of the four sub-figure reflects the finite-N nature of the toy model; it is not intended to represent a continuum limit, but to illustrate the qualitative patterns. Fig. 2-4 are represent 10 uniformly spaced time snapshots between $t=0$ and $t=T$, shown with a purple–yellow gradient indicating time evolution.}
\end{figure}

Fig.2 illustrates the impact of particle numbers ($n = 5, 8, 12, 16$) on the system evolution with $K=0.0191$, the corresponding $L=34.5$. From the figure, one can see that as $n$ increases, the maximum time of stable evolution $T$ decreases. To see longer time evolution of the system, with more accurate shape deformation, we set $n=8$ in investigation of the $K$-influence on the system's evolution. 

The evolutions are performed with a fourth-order Hermite predictor-corrector integrator \cite{Makino1992}, 
which explicitly uses both acceleration and jerk for higher stability when particles approach each other. 
The adaptive step-size control, which halves the step when the jerk becomes too large, further ensures numerical stability and physical consistency.
For the representative case $n = 8$, $t_\mathrm{max} = 1.40$, and initial step size
$\Delta t = 5\times10^{-5}$, the relative energy error is $2.051\times10^{-4}$ in the energy-conservative case
and $2.614\times10^{-2}$ when the GW back–reaction damping is included. The error in the dissipative case is expected and remains within a stable range. Because the particle
distance decreases with increasing $n$, the inverse square gravitational force amplifies
local interactions, leading to faster numerical divergence. Thus, the loss of precision
at large $n$ primarily reflects the singular nature of the Newtonian gravity rather than
a limitation of the integration scheme.

We take $K=0.0123, 0.0191, 0.0286$, which corresponds to $L=49.0, 34.5, 25.0$. Fig.3 shows the simulation results. Since $L >> 1$, the particles within each ring undergo rapid inward contraction due to self-gravitational attraction. We select the time window prior to the onset of chaotic behavior. A larger $K$ leads to a smaller initial angular velocity, causing a smaller angular displacement over the same period. The non-uniform gravitational radiation still compresses the system, so that outer particles become denser than inner ones, leading to the banana-shape deformation. In particular, due to the internal gravity, the "bananas" become asymmetric.

Although Newton gravitation cannot handle physics in the strongly-curved spacetime region, these asymmetric deformations motivate us to incorporate relativistic time-dilation effects, which will be introduced in Section IV.

\begin{figure}[H]
\includegraphics[width=0.5\textwidth]{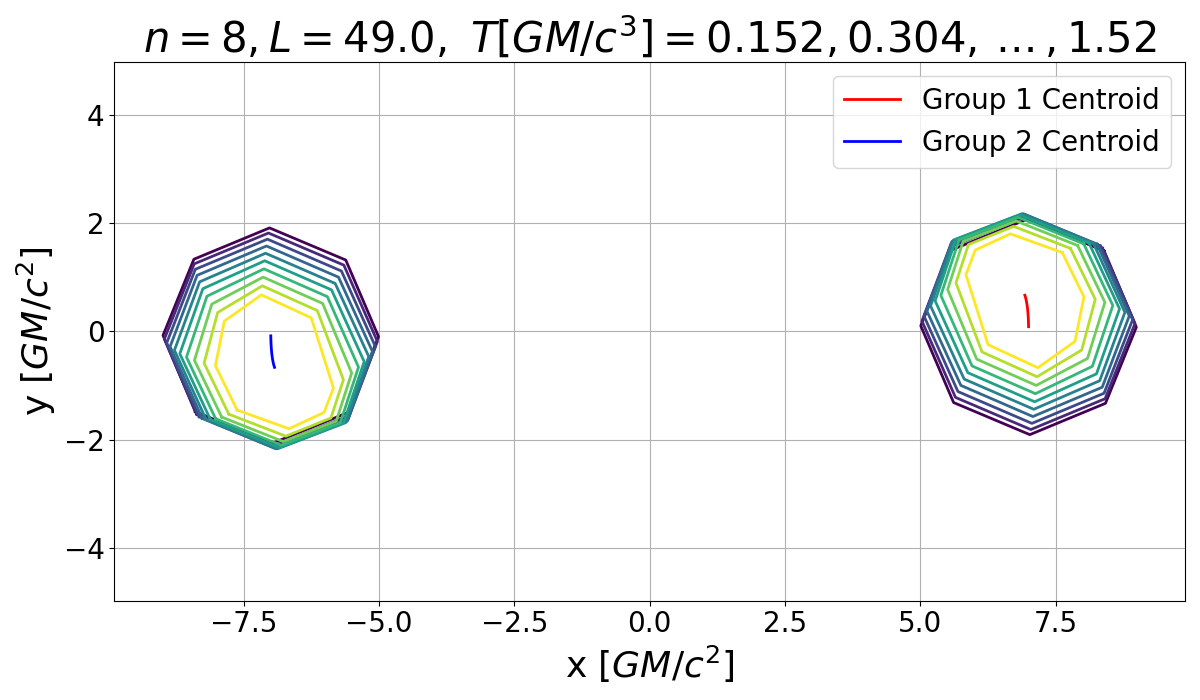} 
\includegraphics[width=0.5\textwidth]{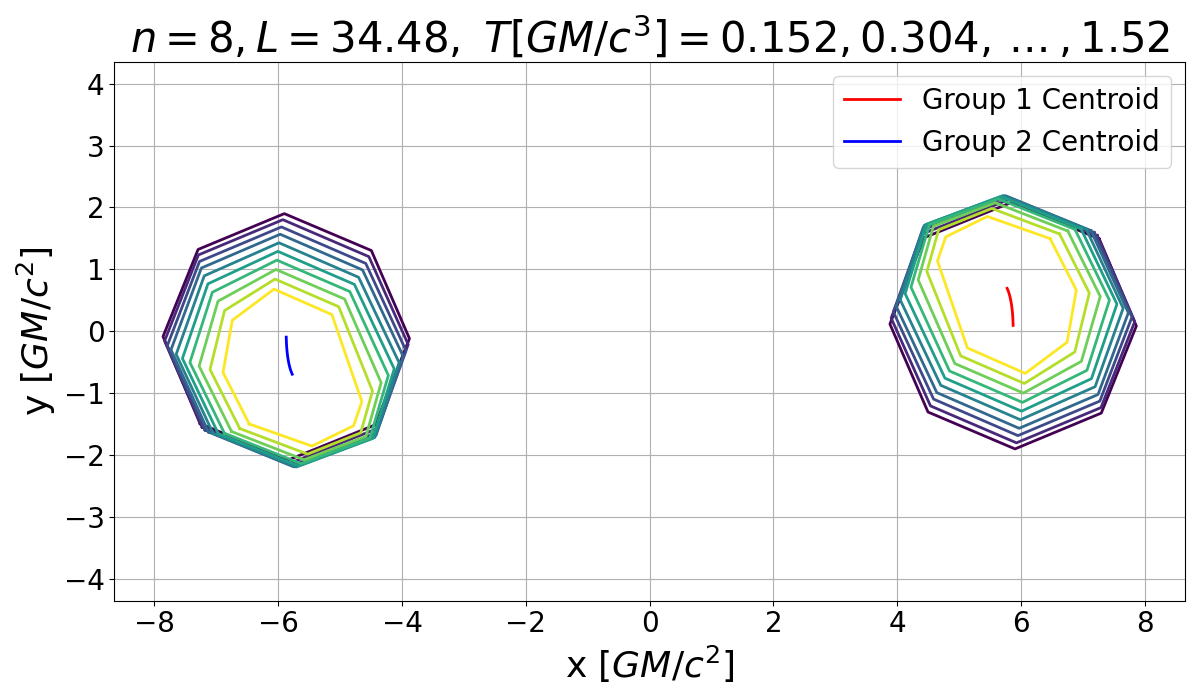} 
\includegraphics[width=0.5\textwidth]{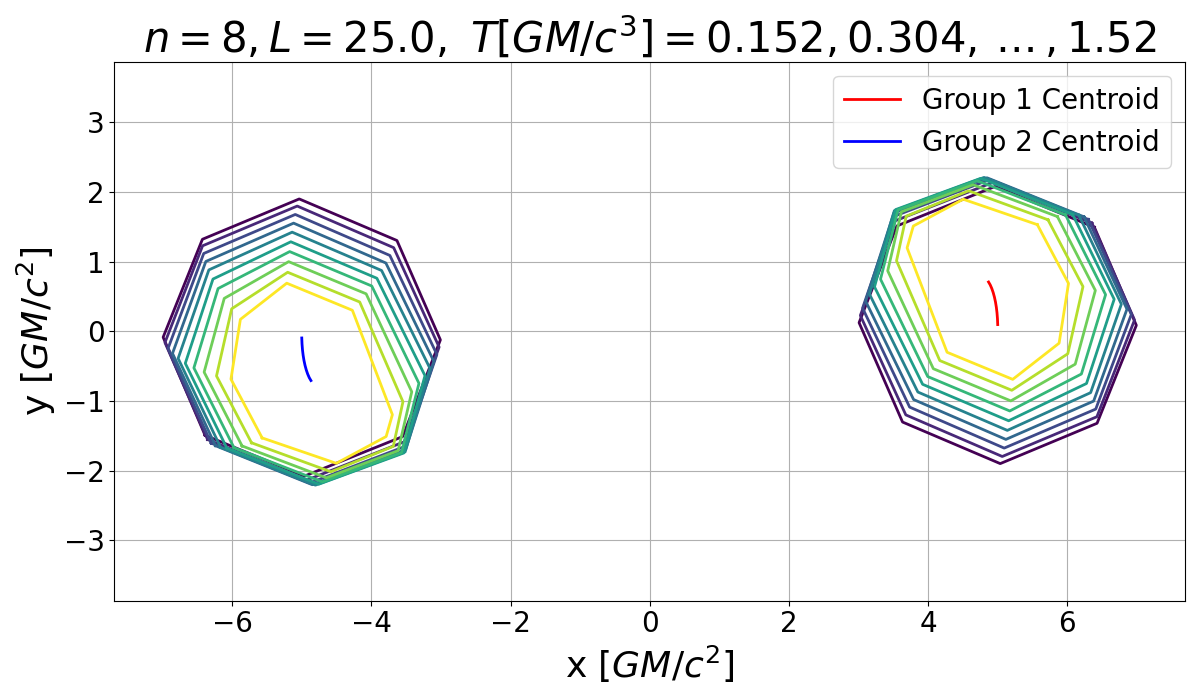} 
\caption{Time evolution of the ring configuration for varying values of $L$ with fixed particle number $n=8$. Asymmetric banana-shape deformations emerge due to uneven GW back-reaction and internal gravitational interactions.} 
\end{figure}

\section{Incorporating Time Dilation Factor}

The Newtonian mechanics framework provides useful qualitative conclusions. To further understand the NR results of BHB, in this section, we introduce GR effects. For a non-rotating, uncharged point mass, the external metric is the Schwarzschild metric:
\beq{}
\mathrm{d}s^2=-\left(1-\frac{2GM}{c^2r}\right)c^2\mathrm{d}t^2+\left(1-\frac{2GM}{c^2r}\right)^{-1}dr^2+r^2\mathrm{d}\theta^2+r^2\sin^2{\theta}\mathrm{d}\phi^2. \label{Schwarz}
\eeq
We will neglect the gravitation propagating time between different particles. This is reasonable for motions around the matter occupation region, because the typical size of such a region is far less than the gravitational wave length, $r\ll\frac{2\pi c}{\sqrt{GM/r^3}}$. Under this condition, the curved spacetime effects can be accounted for through the time dilation factor $\mathrm{d}\tau=\sqrt{-g_{00}}\mathrm{d}t$.
For each particle consisting of the the ring, the time dilation factor reads
\begin{equation}
\gamma_i \mathrm{d}\tau=\mathrm{d}t_i, \quad \gamma_i \equiv \frac{1}{\sqrt{1-\frac{2GM_i}{c^2r_i}}}. 
\label{gammai}
\end{equation}
where $GM_i/r_i=U_i(r_i)$ represents the total Newtonian gravitational potential experienced by particle $i$ due to all the other particles.
This factor is a manifestation of the GR time dilation effects, not a modification of GR itself. Our doing is similar to the idea of looking Newtonian gravitation as a warped-time effects \cite{PoissonWill2014}
\begin{equation}
\mathrm{d}s^{2} = -\left[1 - \frac{2U(r)}{c^2}\right]c^2\mathrm{d}t^2+\mathrm{d}r^2+r^2\mathrm{d}\theta^2+r^2\sin^2{\theta}\mathrm{d}\phi^2,
\end{equation}
but cannot be justified or exactly derived out from the weak-field expansion of GR. We added the time dilation factor \eqref{gammai} into eqs.\eqref{Econ}-\eqref{Lconn} to prohibit the particles consisting of the ring falling into the effective event horizon of the system defined by $1-\frac{2GM_i}{c^2 r_i}=0$. Since this effective horizon is neither the event horizon nor apparent horizon of NR, we do not expect our results to coincide with NR exactly. We only expect qualitative similarities.

In the numerical simulation, the time dilation factor is applied to each time step. The system's initial setup remains the same as that in \eqref{ini.a}, \eqref{ini.b} and \eqref{KeplerFormula}, with $n=128$, as the time-dilation factor suppresses the rapid collapse–induced divergence. 
As the Schwarzschild coordinates become ill-defined at the event horizon, each ring’s mass distribution is initialized at a radial position slightly outside the Schwarzschild radius, i.e.,
\begin{equation}
a_0^\prime=(1+\epsilon)a_0, \quad \epsilon \sim 10^{-3}. \label{aprime}
\end{equation}
Other initial conditions remain the same as those in section \ref{secInternalStructure}. Our numeric simulation results are shown in Fig.4. The RK45 integration method is employed in the simulation displayed in this figure. The maximum residual error over all time-steps is $1.016\times10^{-5}$.

\begin{figure}[ht]
	\includegraphics[width=1.0\textwidth]{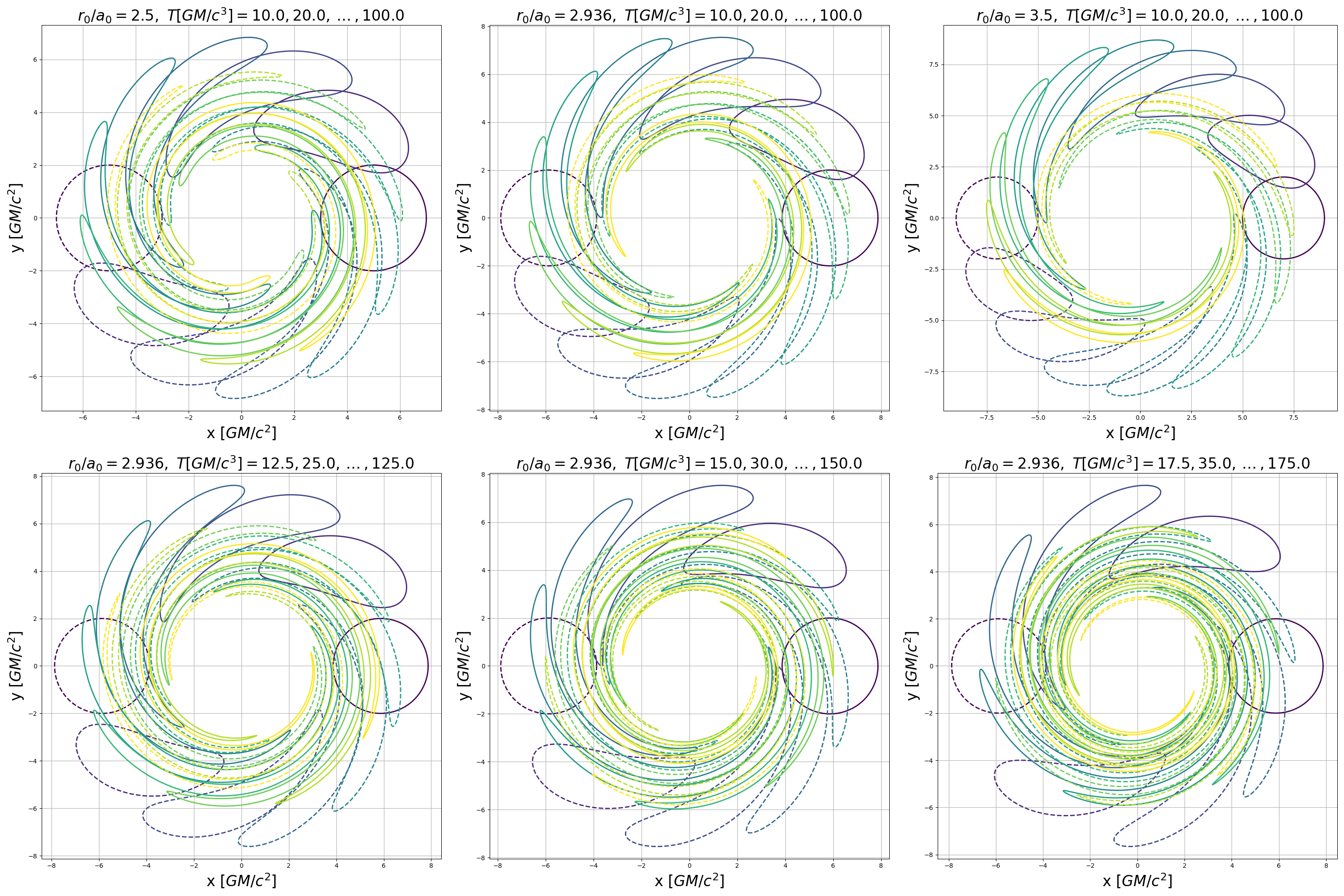} 
	\caption{Deformation of the system for varying initial radius ratios $r_0/a_0$ and evolution times $T$. Top row: comparison of different $r_0/a_0$ at fixed times, showing stronger orbit contraction for smaller ratios. Bottom row: time evolution at fixed $r_0/a_0$, where the system gradually contracts into a spiral structure before stabilizing.} 
\end{figure}

With the introduction of the time dilation factor, the system does not undergo rapid collapse. Fig.4 shows that in the early stages, the ring structure remains consistent with the results from the previous section’s simulation and NR studies \cite{Centrella2010}. As the evolution progresses, the deformation intensifies, forming a rotating, multi-layered spiral structure. This structure resembles a typhoon, with centrally symmetric "spiral arms" on the outer region and a "typhoon eye" in the center. Our results are consistent with the toroidal topology characteristics of NR \cite{Toroidal99a,Toroidal99b,Toroidal16,Toroidal12}.

We hypothesize that, after undergoing internal oscillations, the merged object will gradually fill the "typhoon eye" with mass, while the "spiral arms" lose energy due to quadrupole radiation, leading to their gradual contraction and disappearance. Eventually, the system approaches a more rotationally symmetric, Kerr-like state. This process provides new insights into the ringdown stage of BHB mergers and enhances our understanding of its underlying physical mechanisms.

\section{The Gravitational Waveforms with XOB}
In the previous sections, we have provided an intuitive description for the merger dynamics of the rings. Next, we will calculate the GW-like signal under this scenario and compare the characteristics with the NR waveform. And we use the XOB method \cite{dfzeng2024}, which is capable of describing BHB systems with extended internal structures and constructing GW-like signal.

The Lagrangian of the binary system reads: 
\begin{equation}
\mathcal{L}=-M_A c \sqrt{-g_{\mu\nu}^B \dot{x}_A^\mu \dot{x}_A^\nu}-M_B c \sqrt{-g_{\mu\nu}^A \dot{x}_B^\mu \dot{x}_B^\nu}+\mathcal{L}_{\mathrm{diss}},
\end{equation}
where $\mathcal{L}_{\mathrm{diss}}$ describes the energy dissipation due to GW radiation. Using Legendre transformation to gain the Hamiltonian, in the equal mass case, the orbital Hamiltonian is:
\begin{equation}
H(r)=\frac{2M c^2(1-\frac{GM}{2c^2r})}{\sqrt{1-\frac{3GM}{4c^2r}}}.
\label{Ham}
\end{equation}
The orbital Hamiltonian is dissipated due to gravitational radiation, which can be written as:
\begin{equation}
\frac{\mathrm{d}H}{\mathrm{d}t}=\frac{\mathrm{d}H}{\mathrm{d}r}\frac{\mathrm{d}r}{\mathrm{d}t}=P_{\mathrm{diss}}=-\frac{128G}{5c^5}M^2r^4\omega^6\zeta^2,
\label{Fdiss}
\end{equation}
where $\zeta(t)=\sin[z(t)]/z(t)$ is the radiation activity factor of the system, whose form is exactly derivable from purely geometric calculations. $z(t)$ is the stretching\&bending factor of the rings under the radiation back-reactions. They describes that, as the system evolves, its rotational symmetry becomes increasingly enhanced, thereby reducing the radiation intensity.

We chose the initial data with $r_0/a_0=2.936$ from the previous section, and define the ratio of stetching\&bending as:
\begin{equation}
z(t)=\frac{\text{circumference of the ring on time }t}{\text{prime circumference of the ring}}.
\end{equation}
We fit the data to characterize the ring’s stretching and bending degree. During the early inspiral phase, the gravitational radiation power increases gradually, enhancing the rate of deformation and accelerating the merger process. However, as the black holes approach each other, relativistic time-dilation effects become increasingly significant, slowing down the deformation rate despite continued energy loss. To accurately capture this behavior, we adopt a hyperbolic tangent function. This functional form reflects the initial rapid growth followed by saturation, which aligns well with the qualitative trend observed in our simulations. Fig.5 shows the obtained fitting results. Its expression is given by
\begin{equation}
z(t) = 7.794 \tanh\left( 0.09256 (t/2.4 - 63.06) \right) + 5.766.
\label{fit}
\end{equation}

\begin{figure}[ht]
\centering
\includegraphics[width=0.8\textwidth]{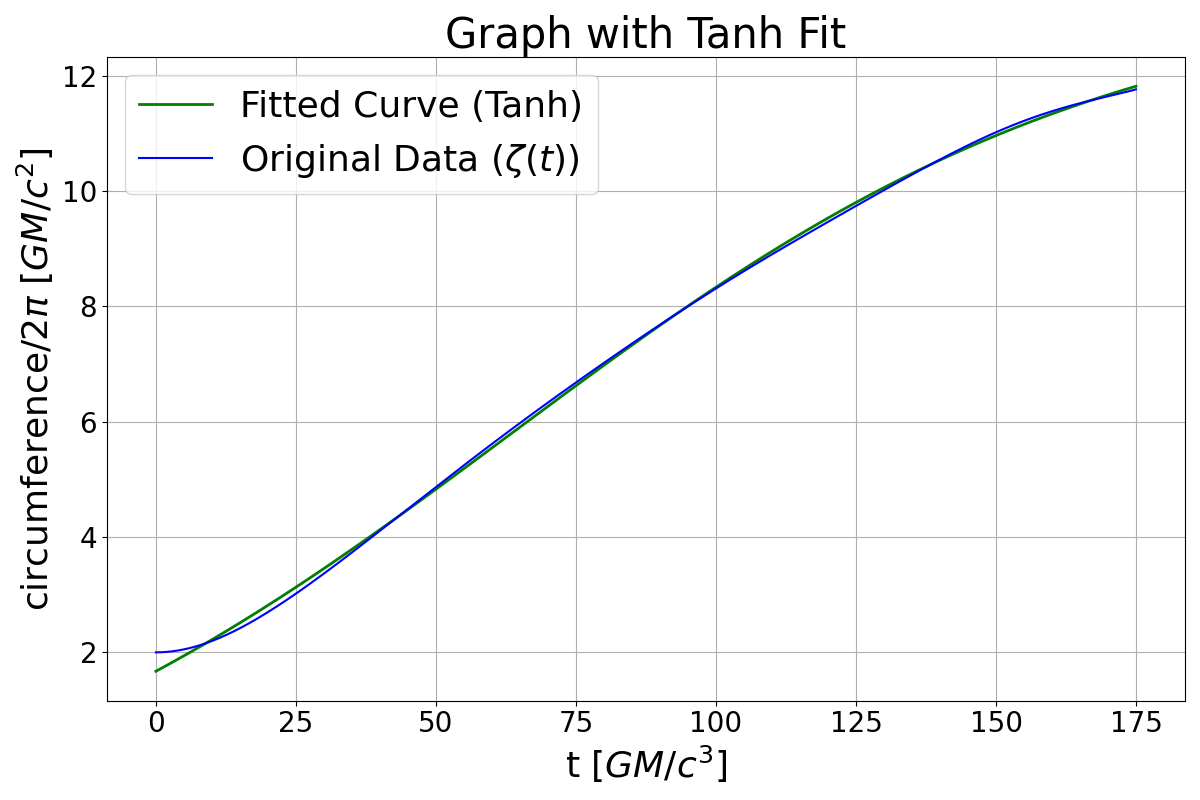}  
\caption{Fitting results of circumference of the ring over time. The coefficient of determination is $R^2=0.999549$.}  
\end{figure}

The strain $h(t)$ is written as:
\begin{equation}
h(t) \propto M_{\mathrm{tot}} r_c^2 \omega^2 \zeta^2(t) \cos\left( 2 \int \omega \, \mathrm{d}t \right),
\end{equation}
where $M_{\text{tot}}$ represents the total mass of the system, $r_c$ is the center of the mass of the ring, $\omega$ is the angular velocity of the ring, $\zeta(t)$ is the deformation factor. Fig.6 shows our waveform results. 

\begin{figure}[ht]
\centering
\includegraphics[width=0.87\textwidth]{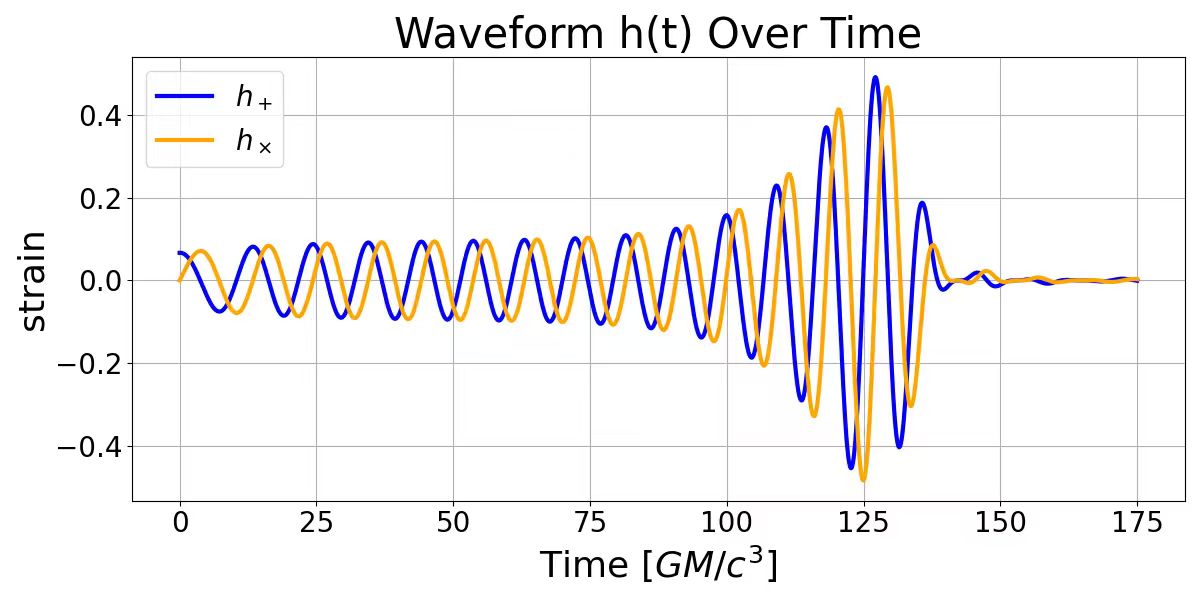}
\includegraphics[width=0.9\textwidth]{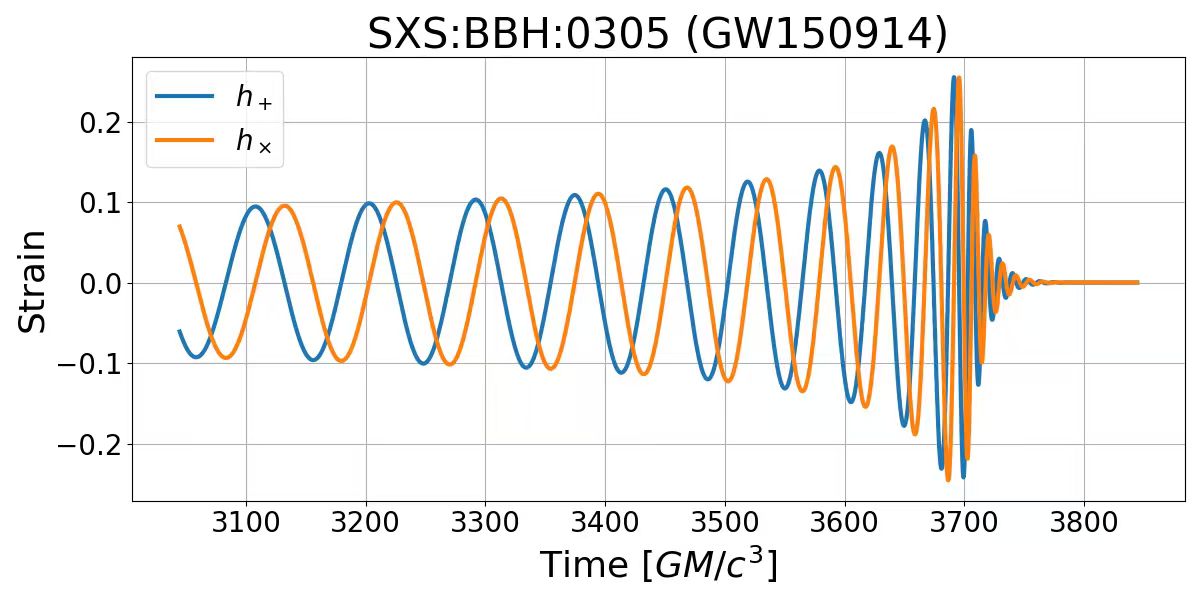}
\caption{The GW-like signal calculated by XOB (upper) and the GW150914 waveform from SXS (lower).}  
\end{figure}

The qualitative agreement between our waveform and NR simulation is evident: the amplitude and frequency increase during the inspiral, reach a peak near merger, and then enter the ring-down stage.
The late-time waveform has a damped envelope, but the small residual oscillation persists at late times, resembling a GW echo.

This feature arises because, the Hamiltonian (\ref{Ham}) of our model has a local minimum at $r=1.5GM$. So the system is dynamically driven toward this minimum rather than undergoing collapse.
For the ring elements at $r>1.5GM$, GW backreaction drives orbital decay. Once any part of a ring crosses inside $r=1.5GM$, the ring’s self-gravity (with time-dilation effects) pulls it back toward $\sim 1.5GM$, producing back-and-forth oscillations.
Each crossing transiently changes the quadrupole moment, generating late-time echo-like bursts in the strain $h(t)$.
The echoes can thus be interpreted as an effective near-horizon reflection enabled by internal structure, in qualitative analogy with echo models \cite{Cardoso2016,Mark2017,Cardoso2018}.

These results suggest that the non-uniform evolution of the internal mass distribution leaves detectable imprints on GW signals, helps us gain a more intuitive understanding of the merger dynamics of BHB.

\section{Conclusion}

Using Newton gravitation and supplements motivated by general relativity, we modeled in this work the in-orbital-plane dynamics of the black hole binary merger process with two co-planar particle rings in three successive steps.

Firstly, assuming a uniform angular velocity and neglecting the self-gravity within each ring, we showed that inhomogeneous GW emission naturally induces a banana-shape deformation.  
Secondly, upon introducing internal gravitation and allowing individual particles to evolve with independent angular velocities, we observe that the deformation become increasingly asymmetric. However, strong self-gravity precipitates rapid collapse and limits the numerically tractable timespan.  

Thirdly, by introducing a time dilation factor inspired by GR, we  follow the system deep into the merger and ringdown phases. The system winds into a typhoon-like spiral whose geometry presents a toroidal topology.  We quantify the banana-shape deformation of the ring by its circumference  and fit the deformation factor as a time-dependent function. Utilizing this fitted function and the XOB approach, we calculate the GW waveform of the process. The results exhibit qualitative features observed in full NR, especially the late-time damping feature.

These findings demonstrate that a Newtonian framework, when judiciously augmented with relativistic corrections, can recover both qualitative morphology and waveform characteristics of BHB merger dynamics. This provides a clear and intuitive perspective on physics behind NR. In particular, our results underscore the possibility that GW waveforms encode information about the internal structure of merging black holes.

Our doing can be naturally extended to the unequal-mass and spinning binaries to uncover more physics between the GW waveform and the inner-structure of merger bodies. However, we must be clear that to hunt for high precision or high matching with Numerical relativity is of nonsense. This kind of works help to identify the existence of relations between the GW waveform and the merger bodies,  but cannot calculate them as precise as numerical relativity. Accepting the existence of this relation, reconstructing the internal structure of BHs using accurately detected GW waveforms will also be a valuable work.

\bmhead{Acknowledgements}
	
This work is supported by the NSFC grant no. {11875082}. 


\begin{thebibliography}{13}
		\ifx \bisbn   \undefined \def \bisbn  #1{ISBN #1}\fi
		\ifx \binits  \undefined \def \binits#1{#1}\fi
		\ifx \bauthor  \undefined \def \bauthor#1{#1}\fi
		\ifx \batitle  \undefined \def \batitle#1{#1}\fi
		\ifx \bjtitle  \undefined \def \bjtitle#1{#1}\fi
		\ifx \bvolume  \undefined \def \bvolume#1{\textbf{#1}}\fi
		\ifx \byear  \undefined \def \byear#1{#1}\fi
		\ifx \bissue  \undefined \def \bissue#1{#1}\fi
		\ifx \bfpage  \undefined \def \bfpage#1{#1}\fi
		\ifx \blpage  \undefined \def \blpage #1{#1}\fi
		\ifx \burl  \undefined \def \burl#1{\textsf{#1}}\fi
		\ifx \doiurl  \undefined \def \doiurl#1{\url{https://doi.org/#1}}\fi
		\ifx \betal  \undefined \def \betal{\textit{et al.}}\fi
		\ifx \binstitute  \undefined \def \binstitute#1{#1}\fi
		\ifx \binstitutionaled  \undefined \def \binstitutionaled#1{#1}\fi
		\ifx \bctitle  \undefined \def \bctitle#1{#1}\fi
		\ifx \beditor  \undefined \def \beditor#1{#1}\fi
		\ifx \bpublisher  \undefined \def \bpublisher#1{#1}\fi
		\ifx \bbtitle  \undefined \def \bbtitle#1{#1}\fi
		\ifx \bedition  \undefined \def \bedition#1{#1}\fi
		\ifx \bseriesno  \undefined \def \bseriesno#1{#1}\fi
		\ifx \blocation  \undefined \def \blocation#1{#1}\fi
		\ifx \bsertitle  \undefined \def \bsertitle#1{#1}\fi
		\ifx \bsnm \undefined \def \bsnm#1{#1}\fi
		\ifx \bsuffix \undefined \def \bsuffix#1{#1}\fi
		\ifx \bparticle \undefined \def \bparticle#1{#1}\fi
		\ifx \barticle \undefined \def \barticle#1{#1}\fi
		\bibcommenthead
		\ifx \bconfdate \undefined \def \bconfdate #1{#1}\fi
		\ifx \botherref \undefined \def \botherref #1{#1}\fi
		\ifx \url \undefined \def \url#1{\textsf{#1}}\fi
		\ifx \bchapter \undefined \def \bchapter#1{#1}\fi
		\ifx \bbook \undefined \def \bbook#1{#1}\fi
		\ifx \bcomment \undefined \def \bcomment#1{#1}\fi
		\ifx \oauthor \undefined \def \oauthor#1{#1}\fi
		\ifx \citeauthoryear \undefined \def \citeauthoryear#1{#1}\fi
		\ifx \endbibitem  \undefined \def \endbibitem {}\fi
		\ifx \bconflocation  \undefined \def \bconflocation#1{#1}\fi
		\ifx \arxivurl  \undefined \def \arxivurl#1{\textsf{#1}}\fi
		\csname PreBibitemsHook\endcsname
		
		\bibitem{Hawking1972} 
		Hawking, S.W. (1972). Black holes in general relativity. \textit{Commun. Math. Phys.}, 25, 152–166.
		
		\bibitem{Bejarano2017} 
		Bejarano, C., Olmo, G.J., \& Rubiera-Garcia, D. (2017). What is a singular black hole beyond general relativity? \textit{Phys. Rev. D}, 95(6), 064043.
		
		\bibitem{HawkingPenrose1970} 
		Hawking, S.W., \& Penrose, R. (1970). The Singularities of Gravitational Collapse and Cosmology. \textit{Proc. R. Soc. Lond. A}, 314(1519), 529–548.
		
		\bibitem{Hawking1974} 
		Hawking, S.W. (1974). Black hole explosions? \textit{Nature}, 248, 30–31.
		
		\bibitem{Almeida2021} 
		Almeida, C.R. (2021). The thermodynamics of black holes: from Penrose process to Hawking radiation. \textit{Eur. Phys. J. H}, 46(1), 20.
		
		\bibitem{Wallace2018} 
		Wallace, D. (2018). The case for black hole thermodynamics part I: Phenomenological thermodynamics. \textit{Stud. Hist. Phil. Mod. Phys.}, 64, 78–88.
		
		\bibitem{Danielsson1993} 
		Danielsson, U.H., \& Schiffer, M. (1993). Quantum mechanics, common sense, and the black hole information paradox. \textit{Phys. Rev. D}, 48(10), 4779–4784.
		
		\bibitem{Hossenfelder2010} 
		Hossenfelder, S., \& Smolin, L. (2010). Conservative solutions to the black hole information problem. \textit{Phys. Rev. D}, 81(6), 064009.
		
		\bibitem{Witten1991} 
		Witten, E. (1991). String theory and black holes. \textit{Phys. Rev. D}, 44(2), 314–324.
		
		\bibitem{Assanioussi2020} 
		Assanioussi, M., Dapor, A., \& Liegener, K. (2020). Perspectives on the dynamics in a loop quantum gravity effective description of black hole interiors. \textit{Phys. Rev. D}, 101(2), 026002.
		
		\bibitem{Modesto2006} 
		Modesto, L. (2006). Loop quantum black hole. \textit{Class. Quantum Gravity}, 23(18), 5587–5601.
		
		\bibitem{Bojowald2020} 
		Bojowald, M. (2020). Black-Hole Models in Loop Quantum Gravity. \textit{Universe}, 6(8), 125.
		
		\bibitem{Cadoni2022} 
		Cadoni, M., Oi, M., \& Sanna, A.P. (2022). Effective models of nonsingular quantum black holes. \textit{Phys. Rev. D}, 106(2), 024030.
		
		\bibitem{Frolov2016} 
		Frolov, V.P. (2016). Notes on nonsingular models of black holes. \textit{Phys. Rev. D}, 94(10), 104056.
		
		\bibitem{Almheiri2013} 
		Almheiri, A., Marolf, D., Polchinski, J., et al. (2013). Black holes: complementarity or firewalls? \textit{J. High Energy Phys.}, 2013(2), 62.
		
		\bibitem{Mathur2014} 
		Mathur, S.D., \& Turton, D. (2014). The flaw in the firewall argument. \textit{Nucl. Phys. B}, 884, 566–611.
		
		\bibitem{Nielsen2009} 
		Nielsen, A.B. (2009). Black holes and black hole thermodynamics without event horizons. \textit{Gen. Relativ. Gravit.}, 41(7), 1539–1584.
		
		\bibitem{Murata2013} 
		Murata, K., Reall, H.S., \& Tanahashi, N. (2013). What happens at the horizon(s) of an extreme black hole? \textit{Class. Quantum Gravity}, 30(23), 235007.
		
		\bibitem{Abramowicz2002} 
		Abramowicz, M.A., Klu{\.z}niak, W., \& Lasota, J.-P. (2002). No observational proof of the black-hole event-horizon. \textit{Astron. Astrophys.}, 396(3), L31–L34.
		
		\bibitem{Frolov2014} 
		Frolov, V.P. (2014). Information loss problem and a ‘black hole’ model with a closed apparent horizon. \textit{J. High Energy Phys.}, 2014(5), 49.
		
		\bibitem{Thornburg1996} 
		Thornburg, J. (1996). Finding apparent horizons in numerical relativity. \textit{Phys. Rev. D}, 54(8), 4899–4918.
		
		\bibitem{Schnetter2006} 
		Schnetter, E., Krishnan, B., \& Beyer, F. (2006). Introduction to dynamical horizons in numerical relativity. \textit{Phys. Rev. D}, 74(2), 024028.
		
		\bibitem{Centrella2010} 
		Centrella, J., Baker, J.G., Kelly, B.J., \& van Meter, J.R. (2010). Black-hole binaries, gravitational waves, and numerical relativity. \textit{Rev. Mod. Phys.}, 82(4), 3069–3119.
		
		\bibitem{dfzeng2021} 
		Zeng, D. (2022). Spontaneous Radiation of BHs. \textit{Nucl. Phys. B}, 977, 115722.
		
		\bibitem{dfzeng2022} 
		Zeng, D. (2023). Gravity Induced Spontaneous Radiation. \textit{Nucl. Phys. B}, 990, 116171.
		
		\bibitem{dfzeng2020} 
		Zeng, D. (2020). Exact inner metric and microscopic state of AdS$_3$-Schwarzschild BHs. \textit{Nucl. Phys. B}, 954, 115001.
		
		\bibitem{dfzeng2018a} 
		Zeng, D. (2018). Schwarzschild Fuzzball and Explicitly Unitary Hawking Radiations. \textit{Nucl. Phys. B}, 930, 533–544.
		
		\bibitem{dfzeng2018b} 
		Zeng, D. (2018). Information missing puzzle, where is hawking's error? \textit{Nucl. Phys. B}, 941, 665–684.
		
		\bibitem{dfzeng2017} 
		Zeng, D. (2017). Resolving the Schwarzschild singularity in both classic and quantum gravities. \textit{Nucl. Phys. B}, 917, 178–208.
		
		\bibitem{dfzeng2024} 
		Zeng, D. (2024). Microscopic state of BHs and an exact one body method for binary dynamics in general relativity. \textit{Eur. Phys. J. C}, 84(4), 370.
		
		\bibitem{Buonanno1999} 
		Buonanno, A., \& Damour, T. (1999). Effective one-body approach to general relativistic two-body dynamics. \textit{Phys. Rev. D}, 59(8), 084006.
		
		\bibitem{Damour2007} 
		Damour, T., \& Nagar, A. (2007). Faithful effective-one-body waveforms of small-mass-ratio coalescing black hole binaries. \textit{Phys. Rev. D}, 76(6), 064028.
		
		\bibitem{GW150914} 
		Abbott, B.P., et al. [LIGO Scientific Collaboration and Virgo Collaboration] (2016). Observation of Gravitational Waves from a Binary Black Hole Merger. \textit{Phys. Rev. Lett.}, 116(6), 061102.
		
		\bibitem{Abbott2016GW2} 
		Abbott, B.P., et al. [LIGO Scientific Collaboration and Virgo Collaboration] (2016). GW151226: Observation of Gravitational Waves from a 22-Solar-Mass Binary Black Hole Coalescence. \textit{Phys. Rev. Lett.}, 116(24), 241103.
		
		\bibitem{Abbott2017GW170104} 
		Abbott, B.P., et al. [LIGO Scientific Collaboration and Virgo Collaboration] (2017). GW170104: Observation of a 50-Solar-Mass Binary Black Hole Coalescence at Redshift 0.2. \textit{Phys. Rev. Lett.}, 118(22), 221101.
		
		\bibitem{Abbott2017GW170814} 
		Abbott, B.P., et al. [LIGO Scientific Collaboration and Virgo Collaboration] (2017). GW170814: A Three-Detector Observation of Gravitational Waves from a Binary Black Hole Coalescence. \textit{Phys. Rev. Lett.}, 119(14), 141101.
		
		\bibitem{Abbott2020GW190412} 
		Abbott, R., et al. [LIGO Scientific Collaboration and Virgo Collaboration] (2020). GW190412: Observation of a binary-black-hole coalescence with asymmetric masses. \textit{Phys. Rev. D}, 102(4), 043015.
		
		\bibitem{Abbott2020GW190521} 
		Abbott, R., et al. [LIGO Scientific Collaboration and Virgo Collaboration] (2020). GW190521: A Binary Black Hole Merger with a Total Mass of 150 $M_\odot$. \textit{Phys. Rev. Lett.}, 125(10), 101102.
		
		\bibitem{Toroidal99a} 
		Husa, S., \& Winicour, J. (1999). Asymmetric merger of black holes. \textit{Phys. Rev. D}, 60(8), 084019.
		
		\bibitem{Toroidal99b} 
		Winicour, J. (1999). The Characteristic Treatment of Black Holes. \textit{Prog. Theor. Phys. Suppl.}, 136, 57–71.
		
		\bibitem{Toroidal16} 
		Bohn, A., Kidder, L.E., \& Teukolsky, S.A. (2016). Toroidal horizons in binary black hole mergers. \textit{Phys. Rev. D}, 94(6), 064009.
		
		\bibitem{Toroidal12} 
		Cohen, M.I., Kaplan, J.D., \& Scheel, M.A. (2012). Toroidal horizons in binary black hole inspirals. \textit{Phys. Rev. D}, 85(2), 024031.
		
		\bibitem{PoissonWill2014} 
		Poisson, E., \& Will, C.M. (2014). \textit{Gravity: Newtonian, Post-Newtonian, Relativistic}. Cambridge University Press.
		
		\bibitem{Peters1963} 
		Peters, P.C., \& Mathews, J. (1963). Gravitational radiation from point masses in a Keplerian orbit. \textit{Phys. Rev.}, 131(1), 435–440.
		
		\bibitem{Wagoner1976} 
		Wagoner, R.V., \& Will, C.M. (1976). Post-Newtonian gravitational radiation from orbiting point masses. \textit{Astrophys. J.}, 210, 764–775.
		
		\bibitem{Turner1977} 
		Turner, M.S. (1977). Gravitational radiation from point masses in unbound orbits: Newtonian results. \textit{Astrophys. J.}, 216, 610–619.
		
		\bibitem{Makino1992} 
		Makino, J., \& Aarseth, S.J. (1992). On a Hermite Integrator with Ahmad-Cohen Scheme for Gravitational Many-Body Problems. \textit{Publ. Astron. Soc. Jpn.}, 44(2), 141–151.
		
		\bibitem{Cardoso2016} 
		Cardoso, V., Hopper, S., Macedo, C.F.B., Palenzuela, C., \& Pani, P. (2016). Gravitational-wave signatures of exotic compact objects and of quantum corrections at the horizon scale. \textit{Phys. Rev. D}, 94(8), 084031.
		
		\bibitem{Mark2017} 
		Mark, Z., Zimmerman, A., Du, S.M., \& Chen, Y. (2017). A recipe for echoes from exotic compact objects. \textit{Phys. Rev. D}, 96(8), 084002.
		
		\bibitem{Cardoso2018} 
		Cardoso, V., Ikeda, T., Moore, C.J., \& Pani, P. (2018). Gravitational-wave echoes and their impact on searches for compact binaries. \textit{Phys. Rev. D}, 98(4), 044018.
		
		
		
	\end{thebibliography}
\end{document}